\documentstyle[12pt]{article}

\def\thefootnote{\fnsymbol{footnote}}

\def\non{\nonumber}
\def\lab{\label}
\def\bib{\bibitem}

\def\gapone{\mbox{\hspace{1cm}}}

\def\bea {\begin{eqnarray}}
\def\eea {\end{eqnarray}}
\def\be {\begin{equation}}
\def\ee {\end{equation}}

\def\dag{\dagger}

\def\ra{\rightarrow}
\def\l{\left}
\def\r{\right}

\def\del{\partial}

\def\lsim{\mbox{{\scriptsize \raisebox{-.9ex}
      {$\;\stackrel{{\textstyle <}}{\sim}\,$} }} }
\def\gsim{\mbox{{\scriptsize \raisebox{-.9ex}
      {$\;\stackrel{{\textstyle >}}{\sim}\,$} }} }

\def\Nb{\bar{N}}
\def\vslash{\not\!\!\mbox{\large $v$}}
\def\hyphen{{\mbox{-}}}
\def\half{\frac{1}{2}}

\def\rme{{\rm e}}\def\rmi{{\rm i}}

\def\gA{g_{\mbox{\tiny A}}}
\def\sigKN{\sigma_{\mbox{\tiny KN}}}
\def\mN{m_{\mbox{\tiny N}}}
\def\mK{m_{\mbox{\tiny K}}}
\def\mpi{m_\pi}

\def\gSp{\sigma_{\pi N}}
\def\TpN{T_{\pi N}}

\def\cA{{\cal A}}\def\cB{{\cal B}}

\def\cL{{\cal L}}\def\cM{{\cal M}}
\def\cO{{\cal O}}

\def\bfk{{\bf k}}

\def\gam{\gamma}

\def\CPT{{\small $\chi$PT}}
\def\Hb{{\small HB$\chi$PF}}
\def\cLHB{\cL_{\rm HB}}

\def\ie{{\it i.e.\ }}

\def\eg{{\it e.g.}}
\def\etal{{\it et al.}}
\def\ibid{{\it ibid\,\,\,}}

\def\PL{{\it Phys. Lett.\,\,}}
\def\NP{{\it Nucl. Phys.\,\,}}
\def\PR{{\it Phys. Rev.\,\,}}
\def\PRL{{\it Phys. Rev. Lett.\,\,}}

\catcode`\@=11
\long\def\@makefntext#1{
\protect\noindent \hbox to 3.2pt {\hskip-.9pt
$^{{\ninerm\@thefnmark}}$\hfil}#1\hfill}

\def\@makefnmark{\hbox to 0pt{$^{\@thefnmark}$\hss}}

\def\ps@myheadings{\let\@mkboth\@gobbletwo
\def\@oddhead{\hbox{}
\rightmark\hfil\ninerm\thepage}
\def\@oddfoot{}\def\@evenhead{\ninerm\thepage\hfil
\leftmark\hbox{}}\def\@evenfoot{}
\def\sectionmark##1{}\def\subsectionmark##1{}}

\renewcommand{\thefootnote}{\fnsymbol{footnote}}

\newcounter{sectionc}\newcounter{subsectionc}
\newcounter{subsubsectionc}
\renewcommand{\section}[1]
{\vspace*{0.6cm}\addtocounter{sectionc}{1}
\setcounter{subsectionc}{0}
\setcounter{subsubsectionc}{0}\noindent
        {\normalsize\bf\thesectionc. #1}
\par\vspace*{0.4cm}}
\renewcommand{\subsection}[1]
{\vspace*{0.6cm}\addtocounter{subsectionc}{1}
        \setcounter{subsubsectionc}{0}\noindent
        {\normalsize\it\thesectionc.\thesubsectionc.
#1}\par\vspace*{0.4cm}}
\renewcommand{\subsubsection}[1]
{\vspace*{0.6cm}\addtocounter{subsubsectionc}{1}
        \noindent
{\normalsize\rm\thesectionc.\thesubsectionc.
\thesubsubsectionc.
        #1}\par\vspace*{0.4cm}}

\def\abstracts#1{{
\centering{\begin{minipage}{12.2truecm}
\footnotesize\baselineskip=
12pt\noindent
  \centerline{\footnotesize ABSTRACT}\vspace*{0.3cm}
        \parindent=0pt #1
        \end{minipage}}\par}}

\renewenvironment{thebibliography}[1]
        {\begin{list}{\arabic{enumi}.}
        {\usecounter{enumi}\setlength{\parsep}{0pt}
\setlength{\leftmargin 1.25cm}{\rightmargin 0pt}
         \setlength{\itemsep}{0pt} \settowidth
        {\labelwidth}{#1.}\sloppy}}{\end{list}}

\newcounter{itemlistc}
\newcounter{romanlistc}
\newcounter{alphlistc}
\newcounter{arabiclistc}

\newcommand{\fcaption}[1]{
        \refstepcounter{figure}
        \setbox\@tempboxa =
\hbox{\footnotesize Fig.~\thefigure. #1}
        \ifdim \wd\@tempboxa > 6in
           {\begin{center}
        \parbox{6in}{\footnotesize
\baselineskip=12pt Fig.~\thefigure. #1}
            \end{center}}
        \else
             {\begin{center}
             {\footnotesize Fig.~\thefigure. #1}
              \end{center}}
        \fi}

\newcommand{\tcaption}[1]{
        \refstepcounter{table}
        \setbox\@tempboxa =
\hbox{\footnotesize Table~\thetable. #1}
        \ifdim \wd\@tempboxa > 6in
           {\begin{center}
        \parbox{6in}{\footnotesize\baselineskip=
   12pt Table~\thetable.
#1}
            \end{center}}
        \else
             {\begin{center}
             {\footnotesize Table~\thetable. #1}
              \end{center}}
        \fi}

\def\@citex[#1]#2{\if@filesw\immediate\write\@auxout
        {\string\citation{#2}}\fi
\def\@citea{}\@cite{\@for\@citeb:=#2\do
        {\@citea\def\@citea{,}\@ifundefined
        {b@\@citeb}{{\bf ?}\@warning
    {Citation `\@citeb' on page \thepage \space undefined}}
        {\csname b@\@citeb\endcsname}}}{#1}}

\newif\if@cghi
\def\cite{\@cghitrue\@ifnextchar [{\@tempswatrue
        \@citex}{\@tempswafalse\@citex[]}}
\def\citelow{\@cghifalse\@ifnextchar [{\@tempswatrue
        \@citex}{\@tempswafalse\@citex[]}}
\def\@cite#1#2{{$\null^{#1}$\if@tempswa\typeout
        {IJCGA warning: optional citation argument
        ignored: `#2'} \fi}}

 1
 1
 1

\font\ninerm=cmr9

\begin{document}
\begin{flushright}
USC(NT)-95-2
\end{flushright}
\setcounter{footnote}{2}
\centerline{\normalsize\bf NUCLEAR RESPONSES
TO ELECTRO-WEAK PROBES AND}
\baselineskip=16pt
\centerline{\normalsize\bf IN-MEDIUM
CHIRAL PERTURBATION THEORY\footnote{Supported
in part by the National
Science Foundation
(USA), Grant No. PHYS-9310124.}
\footnote{Invited talk at WEIN'95, Osaka, Japan;
June 1995}}
\baselineskip=22pt
\centerline{\footnotesize K. KUBODERA,\,\,\,
F. MYHRER, \,\,\,Th. MEISSNER\, and\, R. MORONES}
\baselineskip=12pt
\centerline{\footnotesize\it Department
of Physics and Astronomy,
University of South Carolina}
\baselineskip=12pt
\centerline{\footnotesize\it Columbia,
South Carolina 29208, USA}

\vspace{0.6cm}
\abstracts{We discuss two topics concerning
the application of chiral perturbation theory
to nuclear physics:
(1) the latest developments in the study of
possible kaon condensation in dense baryonic systems;
(2) nuclear responses to electro-weak probes}

\normalsize\baselineskip=15pt
\setcounter{footnote}{0}
\renewcommand{\thefootnote}{\alph{footnote}}

\vspace{-0.1cm}
\section{Introduction}

\vspace{-0.4cm}
\indent\indent
Chiral perturbation theory (\CPT) offers
a valuable guiding principle
in our attempt to relate nuclear dynamics
to the fundamental QCD.
The concept of chiral counting also gives
a clear perspective in organizing our description of
complicated nuclear dynamics.
Indeed, a new line of nuclear physics
based on \CPT \,
seems to be steadily gaining ground.
In this talk,
after giving a minimal sketch of \CPT,
we present two examples
of the nuclear physics application of \CPT.
We first discuss the latest developments
in the study of possible kaon condensation
in dense matter.
We then describe the use of \CPT\,
in calculating nuclear responses
to electro-weak interaction probes.

The introduction of \CPT \,follows
a generic pattern to define an effective
theory.$\!$\cite{gl84,wei90,bkm95}
Consider the vacuum-to-vacuum amplitude
in QCD in the presence of external fields
\be
\rme^{iZ[v,a,s,p]}=
\int[d{\mbox{\footnotesize$G$}}]
[dq] [d\bar{q}]\,
\rme^{\rmi\int \!d^4\!x\,
\cL(q,{\bar q},G;\,v,a,s,p)}
\lab{eq:ZQCDsource}
\ee
where
$\cL=\cL^0_{\rm QCD}+\bar{q}\gamma^\mu
[v_\mu(x)-\gamma^5 a_\mu(x)]q
-\bar{q}[s(x)-ip(x)]q$.
The external fields,
$v_\mu$, $a_\mu$, $s$ and $p$,
are assigned appropriate SU(3)$\times$SU(3)
transformation properties
to make $\cL$ chiral invariant.
The effective lagrangian that describes
low-energy phenomena of QCD
($E\,\lsim\Lambda_\chi\!\sim$1 GeV)
involves the Goldstone bosons and
is introduced through
\be
\rme^{iZ[v,a,s,p]}=
\int[d\mbox{\footnotesize{$U$}}]\,
\rme^{\rmi\int \!d^4\!x\,
\cL_{\rm eff}(U;\,v,a,s,p)},
\lab{eq:ZLeff}
\ee
where
$U\equiv
\exp(i\sum_{a=1}^8\pi^a\lambda^a/f_\pi)$
with $\pi^a$ the octet pseudo-scalar mesons.
In \CPT\,we expand $\cL_{\rm eff}$
in powers of $\del_\mu/\Lambda_\chi$ and
the quark mass matrix $\cM/\Lambda_\chi$
and, for a given order of expansion,
retain all terms that are consistent with
the symmetries.
In extending this scheme
to the baryon field $N$,
we realize that $\del_0$
acting on $N$ yields $\sim\mN$,
which is not small compared with $\Lambda_\chi$.
The heavy-baryon chiral perturbation formalism
(\Hb) allows us to avoid this difficulty.$\!$\cite{jm91}
Here, instead of the ordinary Dirac field $N$
we work with $B$ defined by
$B(x)\equiv\rme^{\rmi v\cdot x}N(x)$
with $v\sim(1,0,0,0)$,
shifting the energy reference point
from 0 to $\mN$.
If we are only concerned
with small energy-momenta $Q$
around this new origin,
the antibaryon can be ``integrated away".
$\cL_{\rm eff}(B, U;\,v,a,s,p)$
describing this particle-only world
may be defined similarly to Eq. (\ref{eq:ZLeff}).
The corresponding equation of motion
for $B$ may be rewritten as coupled equations
for the large and small components $B_{\pm}$
defined by
$B_{\pm}\equiv P_{\pm}B$ with
$P_{\pm}\equiv (1\pm\!\vslash)/2$.
Eliminating $B_-$ in favor of $B_+$
leads to an equation of motion for $B_+$.
The \Hb \,lagrangian $\cLHB$
is defined as an effective lagrangian
that reproduces the equation of motion
for $B_+$ and $U$.
Since $B_-\propto (Q/\mN)B_+$,
$\cLHB$ involves expansion
in $\del_\mu/\mN$
as well as in $\del_\mu/\Lambda_\chi$
and $\cM/\Lambda$.
We can organize this expansion as
\be
\cLHB=\cL^{(1)}+\cL^{(2)}+\,\cdots\;\,;\;\;\;\;\;\;
\cL^{(\nu)}=\cO(Q^{\nu-1})
\ee
The chiral order index $\nu$ is defined as
$\nu=d+(n/2)-2$,
where $n$ is the number of
fermion lines involved in a vertex,
and $d$ is the number of derivatives
(with $\cM\propto m_\pi^2$ counted
as two derivatives).
The explicit expression relevant
to the meson-baryon sector is\cite{bkm95}
\be
\cLHB\,=\,\bar{B}_+\left[
\cA^{(1)}+\cA^{(2)}
\,+\,(\gam_0\cB^{(1)}\gam_0)
\frac{1}{2\mN}\cB^{(1)}\right]B_++\cO(Q^2),
\lab{eq:LHBAB}
\ee
The leading order term is given in terms of
$u=\sqrt{U}$ and
$S_\mu=i\gam_5\sigma_{\mu\nu}v^\nu/2$
as
\bea
\cA^{(1)}&=&i(v\cdot D)+\gA(u\cdot S)\\
D_\mu&=&\del_\mu+[u^\dag,\del_\mu u]/2
-i\,u^\dag(v_\mu+a_\mu)u/2
-i\,u(v_\mu-a_\mu)u^\dag/2
\eea
The expressions for higher order terms
can be found in Ref.$\!$\cite{bkm95}

Chiral counting can also be applied to
Feynmann diagrams;
the chiral order $D$ of
an irreducible Feynmann diagram
is given by\cite{wei90}
\be
D\,=\,2-\half N_E+2L-2(C-1)+\sum_i\nu_i,
\lab{eq:Dcount}
\ee
where $N_E$ is the number of external fermion lines,
$L$ the number of loops,
$C$ the number of disconnected parts,
and the sum runs over vertices.

\vspace{-0.2cm}
\section{Kaon Condensation in Dense Baryonic Matter}

\vspace{-0.4cm}
\indent \indent
Kaon condensation in dense baryonic matter
has been discussed
by many authors.$\!$\cite{kn86,kub93a}
According to the latest calculation,$\!$\cite{lbmr95}
the critical density $\rho_c$ for kaon condensation
is $\rho_c \approx 4\rho_0$
($\rho_0=$ normal nuclear matter density) and,
with the Brown-Rho scaling\cite{br91} included,
$\rho_c$ can be as low as $2\rho_0$.
Kaon condensation
(as we are interested in here)
is driven by the $s$-wave interactions,
unlike pion condensation
which depends on the $p$-wave interactions.
The strong $s$-wave $K$-$N$ attraction
comes partly from the so-called $\sigma$-term,
which is significantly stronger
for the kaon than for the pion.
Furthermore, the vector-meson exchange contributions
can give rise to strongly attractive s-wave interactions
for some $K$-$N$ channels,
whereas they are either repulsive
or only weakly attractive for the $\pi N$ channels.
These features motivate us
to examine the possibility of s-wave
kaon condensation.
As far as observational consequences are concerned,
a kaon condensate
(like a boson condensate in general)
could enhance significantly neutrino emission
from nascent neutron stars,
cooling them much faster.
Furthermore, the condensate
can drastically soften the equation of state
for collapsing stars.
Brown and Bethe\cite{bb94} argue that
this softening leads to proliferation
of mini blackholes, which resolves
the long-standing puzzle
that the observational value
for the ratio $R\equiv$
[\# of neutron stars]/[\# of supernova events]
is inexplicably low.

Two of the outstanding issues
facing kaon condensation are
the $m^*_N$ effect and
the off-mass-shell effects
(both to be explained below).
We wish to report here the progress we have made
on these issues over the past year.

\vspace{-0.2cm}
\subsection{The $\mN^*$ Effect}

\vspace{-0.5cm}
\indent\indent
Several authors argued
that in-medium nucleon mass reduction
could strongly hinder
kaon condensation.$\!$\cite{kub94,schetal94}
As mentioned above,
the $K$-$N$ $\sigma$-term,
$\sigKN\bar{\psi}\psi\bar{K}K$,
provides a significant part of the $s$-wave attraction.
The $\sigma$-term attraction
in baryonic matter
is (in the mean-field approximation)
proportional to the Lorentz scalar density
$\rho_s \equiv \,<\!\!\bar{\psi}\psi\!\!>$.
The earlier works, however,
used the approximation $\rho_s \sim \rho$,
where $\rho$ is the baryon density,
$\rho \equiv\,<\!\!\bar{\psi}\gamma_0\psi\!\!>$.
This simplifies the calculation considerably,
since $\rho$ is a conserved quantity
that can be specified as an external parameter,
whereas $\rho_s$ is known only
after the whole dynamics is solved.
For a nucleon of effective mass
$\mN^*$ and momentum $\bfk$,
we have
$\bar{u}_\bfk u_\bfk=
[\mN^*/(\mN^{*2}+\bfk^2)^{1/2}]\,
u^\dag_\bfk u_\bfk$,
which suggests
that using $\rho$ instead of $\rho_s$
overestimates the $\sigma$-term contribution
and that this overestimation becomes
more serious for smaller values of $\mN^*$.
Detailed calculations\cite{schetal94}
based on the Walecka model\cite{sw86}
indicate that, for $\mN^*\lsim 0.75\rho_0$,
the effective kaon mass $\mK^*$
does not any longer go down to zero but
levels off as $\rho$ increases,
and $\mK^*(\rho\!\ra\!\infty)\gsim 0.45 \mK$.
For convenience we refer to this feature
as the ``$\mN^*$ effect".
If the $\mN^*$ effect is indeed as strong
as the Walecka model suggests,
there would be no kaon condensation.

Does this argument invalidate
Lee \etal's conclusion\cite{lbmr95}
$\rho_c=(2\!\!\sim\!\!4)\rho_0$ ?
This issue is connected
to the choice of the nucleon field.
The Walecka model uses the original Dirac field.
For systematic chiral counting, however,
it is more advantageous to work with
the heavy baryon field $B_+$,
and this is what Lee \etal\cite{lbmr95} did.
Now, for $B_+$, there is by construction
no distinction between
$\rho_s\equiv\bar{B}_+B_+$
and $\rho\equiv\bar{B}_+\gam_0 B_+$.
In this sense Lee \etal's approach
is free from the conventional approximation
$\rho_s\approx\rho$.
But this is of course not the whole story.
In \Hb \,\,the effects of the $B_-$ responsible
for $\rho_s\neq\rho$
are transformed into the higher order terms in
$1/\mN$ expansion.
So we need to examine how this
$1/\mN$ expansion is handled in practice.
The lowest-order term in \Hb\,
[\ie\,$\cA^{(1)}$ term in $\cL_{\rm HB}$,
Eq. (\ref{eq:LHBAB})]
applies to an infinitely heavy baryon,
and hence the $\mN^*$ effect
is totally absent here.
The next order contribution contains
$\nu=1$ terms in ordinary chiral counting
($\cA^{(2)}$ term) and terms that are first order
in $1/\mN$.
We denote the latter by $\cL_{1/m}$.
$\cL_{1/m}$ consists of
the baryon kinetic energy term
$\cL_{1/m}^B\equiv
\bar{B}_+(-\del_\mu^2/2\mN)B_+$
and the meson-baryon interaction part
$\cL_{1/m}^{\rm int}$.
Now, to understand the calculational scheme
adopted by Lee \etal,
let us rearrange $\cL_{\rm HB}$ as
\bea
\cL_{\rm HB}&=&
\l(
\cL_{\rm HB}
({\rm non}\!-\!{\rm strange\;sector})
+\cL_{\rm HB}({\rm strange\;sector})
\r)
_{\mN\ra\infty}
+\cL_{1/m}+\cdots
\non\\
&=& \l\{
 \cL_{\rm HB}({\rm non}\!-\!{\rm strange})_
{\mN\ra\infty}
+\cL_{1/m}^B
+\cL_{1/m}^{\rm int}
({\rm non}\!-\!{\rm strange})
\r\}
\non\\
&&\gapone +
\l[
\cL_{\rm HB}({\rm strange})
_{\mN\ra\infty}
+\cL_{1/m}^{\rm int}({\rm strange})
\r]
+\cdots\lab{eq:Lsplit}
\eea
We first discuss the non-strange sector
corresponding to the terms in the curly brackets.
In the existing calculations based on \Hb\,
the energy density for the non-strange sector
is taken from nuclear matter calculations of
the Brueckner-Hartree-Fock type.
This effectively incorporates
the $1/\mN$ correction.
In fact, since any realistic nuclear matter
calculation takes account of the change
$\mN\!\ra\!\mN^*$,
the use of
the nuclear matter calculation results
allows us to go beyond
the $1/\mN$ correction.
This is in a sense a welcome feature
but there is a problem too.
In \Hb\, the change $\mN\!\ra\!\mN^*$
arises either from $(1/\mN)^n$ corrections
($n\geq 2$) or from vertices with $\nu\geq 2$,
and we must deal with
a great multitude of possible terms.
By using the nuclear matter results
containing the effective mass change
one is selecting a very particular subset
of the higher order effects,
and at present there is
no clear justification for doing so.
On the other hand, the fact the change
$\mN\!\ra\!\mN^*$ features importantly
in nuclear matter calculation does indicate
that one cannot simply stop at
the first correction term in $1/\mN$ expansion.

We next discuss the strangeness sector,
the terms in the square brackets
in Eq.~(\ref{eq:Lsplit}).
Here we note that $\cA^{(2)}$ terms
contained in
$\cL_{\rm HB}({\rm strange})$ is of the same
chiral order ($\nu=1$)
and that the coefficients appearing in $\cA^{(2)}$
are in fact phenomenologically fixed
in such a manner that observables
for one-meson one-baryon systems be reproduced.
Then the introduction of the $1/\mN$ term
just leads to a readjustment of
these parameters.
Therefore, the $\mN^*$ effect
in the Walecka model would correspond
to terms of $\nu=2$ or higher.
Again, there are many such terms
and, for consistency, one must retain all of them.
The Walecka model represents
a particular choice of a subset,
and it remains to be seen
whether the strong $\mN^*$ effect
suggested by the model survives
a fully consistent treatment.
On the other hand,
no calculations so far done in \Hb\,
go beyond the $1/\mN$ term
in the strangeness sector.
The only exception is
a qualitative remark by Lee \etal\cite{lbmr95}
that a multifermion term such as
$(\bar{B}_+\gam_\mu B_+)
(\bar{B}_+B_+)\bar{K}\del_\mu K$
can lead to a in-medium
($\mN^*$-dependent) modification
of the $K$-$N$ interaction.
This $\mN^*$ effect in fact
enhances the $K$-$N$ attraction
quite in contrast to the $\mN^*$ effect
found in the Walecka model.
Obviously, more systematic treatments
of higher order terms are required
before we can reach a solid conclusion
on the $\mN^*$ effect.

In this connection,
one may worry that
a plethora of multi-fermion vertices
that can participate in dense matter
will spoil the convergence of chiral expansion.
In fact, this does not happen as easily as
one naively expects.
According to Eq. (\ref{eq:Dcount}),
a Feynmann diagrams with a given number
of external lines $N_E$ has
a smaller value of $C$
if it contains vertices
with larger values of $n$,
thus resulting in a higher chiral order index $D$.
So, the actual contributions of vertices
with large fermion numbers
to a Feynmann diagram
are more suppressed than the chiral counting
of individual vertices would indicate.
This implies that we probably need not deal with
a tower of multi-fermion terms
to understand the $\mN^*$ effect
in the framework of \Hb.
There have been interesting attempts
at relating the Walecka model
to \Hb.$\!$\cite{gr94}

\vspace{-0.2cm}
\subsection{Off-Shell-Effects}

\vspace{-0.4cm}
\indent\indent
Since the main points of our discussion here
can be described more conveniently
for the pion than for the kaon,
we shall discuss the pion case.
According to the standard multiple
scattering theory,
the pion-nuclear optical potential,
or pion self-energy, is given by
\be
\Pi=\rho\,t_{\pi A}+ \cdots,\lab{eq:optical}
\ee
where $t_{\pi A}$ is the $t$-matrix
describing pion scattering off
a nucleon in medium,
and the dots represent processes
involving more than a single scatterer.
The pion propagator pertaining to $t_{\pi A}$
is a full A-body nuclear hamiltonian,
not just the single nucleon hamiltonian.
Note that, in order to use $\Pi$
in the determination of
the in-medium dispersion relation for a pion,
we need information on $t_{\pi A}$
for off-shell as well as on-shell kinematics.
In the low-density limit,
we only need retain the $\rho\,t_{\pi A}$ term,
and furthermore we can replace $t_{\pi A}$
with the on-shell $t$-matrix
for free $\pi$-$N$ scattering.

Now, the issue raised
by Yabu \etal$\!$\cite{ynk93} is as follows.
Consider a toy $\pi$-$N$ lagrangian
that contains only the $\sigma$ term\footnote{
This is a highly simplified version of
the Kaplan-Nelson lagrangian.
Although recent calculations\cite{lbmr95,tw95}
take due account of energy-dependent terms of
the same chiral order as the $\sigma$ term,
our points can be explained
without those additional terms.}:
\be
\cL_{1} = \frac{1}{2} \l[ -\phi (\Box+\mpi^2) \phi
+ \frac{\gSp}{f^2} \phi^2 {\Nb N} \r].\lab{eQa}
\ee
For $\cL_{1}$,
the $\pi$-$N$ scattering amplitude
in tree approximation is simply a constant:
$\TpN^{(1)} = \gSp/f^2$.
The corresponding pion effective mass
$\mpi^*$ (in the mean-field approximation) is
$[\mpi^*(1)]^2 =\mpi^2-\rho\,(\gSp/f^2)$.
On the other hand,
the PCAC plus current algebra
gives the forward scattering amplitude
$\TpN^{(2)} =
[(k^2 +(k')^2-\mpi^2)/f^2\mpi^2]\gSp$.
The corresponding $\mpi^*$ is given by
$[\mpi^*(2)]^2 =\mpi^2 \,
[1 +\rho\,(\gSp/\mpi^2f^2)]\cdot
[1 +2\rho\,(\gSp/\mpi^2f^2)]^{-1}$.
Although $\mpi^*(1)$ and $\mpi^*(2)$
are identical for low densities,
they behave very differently for large values of $\rho$.
In particular, $\mpi^*(2) \!\ra \!\mpi/\sqrt{2}$ as
$\rho \!\!\ra \!\!\infty$.
Yabu \etal, who pointed out this discrepancy,
argued that the existing calculational frameworks
did not allow one to resolve this problem.

It behooves to remember here
the following general points:
(i) The formal definition of
$\mpi^*$ is a value of the energy variable
$\omega$ for which
the exact in-medium Green's function
$G_\rho(x;\phi)=
<\!\rho|T\phi(x)\phi(0)|\rho\!>$
develops a pole (for zero momentum);
(ii) For a given lagrangian,
the physical observable $\mpi^*$
should not depend on the definition
of interpolating fields $\phi$;
(iii) Although off-mass-shell $\pi$-$N$ amplitudes
vary for different choices of $\phi$,
this variation should not affect any observables
including $\mpi^*$;
(iv) Although off-shell $\pi$-$N$ amplitudes are
unphysical in the sense of (iii) and also in that
they cannot be observed in $\pi$-$N$ scattering,
they do constitute ingredients of
larger Feynmann diagrams;
(v) The statements (i)$\sim$(iii) hold true
only if the whole calculation is done exactly.
This last point is trivial but
nonetheless worth emphasizing.

Now, within the framework of
the leading order optical potential,
the variance
between $\mpi^*(1)$ and $\mpi^*(2)$
is a direct consequence of the
fact that $\TpN^{(1)}$ and $\TpN^{(2)}$
have different off-shell behaviors.
Referring to the above general statements,
one could ask
whether this is a manifestation of different dynamics,
or just a spurious off-shell effect
that fails to disappear because of
the approximation used.
Yabu \etal\,favored the first possibility,
conjecturing that different treatments of
multi-fermion terms are responsible
for the different behaviors of
$\TpN^{(1)}$ and $\TpN^{(2)}$.
This interpretation, however, was criticized
by Lee \etal$\!$\cite{lbmr95}
and by Thorsson and
Wirzba (TW).$\!$\cite{tw95}
TW show explicitly that,
starting from the same $\cL_{HB}$,
one can derive either of
$\TpN^{(1)}$ and $\TpN^{(2)}$
by adding to $\cL_{\rm HB}$
different pseudoscalar source terms.
This ensures that,
provided one can calculate
$G_\rho(x;\phi)=
<\!\rho|T\phi(x)\phi(0)|\rho\!>$
{\underline{exactly}},
one would get the same $\mpi^*$
regardless of whether one uses
$\TpN^{(1)}$ or $\TpN^{(2)}$.
Beautiful !!
(Please note, however, the underline
attached to ``exactly".)
In practice, we must adopt some approximation,
the crudest and most commonly used approximation
being $\Pi\approx \rho\,t_{\pi\,N}$.
In these approximate calculations,
choice between $\TpN^{(1)}$ and
$\TpN^{(2)}$ does matter,
and TW's formal proof is not of immediate help
in making this choice.

We must mention here, however,
another important point made by TW.
TW demonstrates that,
within the mean-field approximation,
the use of the effective action leads
to the identical dispersion relation for
an in-medium pion regardless of
different choices of the pseudoscalar source.
This is a remarkable result,
but it seems important to
examine to what extent this theorem
is tied to the mean field approximation.
In fact, if TW's result is valid
beyond the mean filed approximation,
that would give a tremendous impact
to the ``standard" multiple scattering formalism.
We would be forced to conclude that
the obvious off-shell dependence
exhibited by the leading term
in the Watson expansion is spurious
(at least for a system the dynamics of which
is strongly constrained by chiral symmetry).
This point deserves a serious investigation
quite apart from the specific problem of
meson condensation.

\section{Nuclear Responses to Electro-Weak Probes}

\vspace{-0.4cm}
\indent \indent
The nuclear hamiltonian is normally taken to be
$H_N = \sum_{i=1}^{A} T_i + \sum_{i,j}^{A} V_{ij}$,
where $T_i$ is the nucleon kinetic energy,
and $V_{ij}$ is the ``realistic" $N$-$N$ potential.
Arriving at $H_N$ starting
from the fundamental QCD description involves:
(i) translating the quark and gluon degrees of freedom
into the effective degrees of freedom of hadrons;
(ii) truncating the Hilbert space of hadrons
down to that of non-relativistic nucleons
interacting via potentials.
The \CPT \,allows us to carry out (i) and (ii)
in a well-defined way,
preserving the basic chiral properties of QCD.
Construction of the realistic $N$-$N$ potentials
based on \CPT \,was described
by Weinberg\cite{wei90}
and by van Kolck \etal\cite{kol92}
These \CPT\, potentials can reproduce
the $N$-$N$ observables almost as satisfactorily
as the conventional boson-exchange potentials
which contain many {\it ad hoc} parameters.

In the truncated nucleonic space,
nuclear responses to external probes
such as electromagnetic and weak currents
involve not only single-nucleonic terms
(= impulse approximation terms)
but also multi-nucleonic contributions
named the exchange currents.
Here again, \CPT \, provides a systematic framework
for organizing exchange-current contributions
according to their chiral
counting orders.$\!$\cite{rho91,pmr93,pmr95}

A problem in testing the exchange currents
in complex nuclei
is that exact solutions
for the $A$-body Schr\"{o}dinger equation
$H_N\Psi= E\Psi$ are hard to obtain
and therefore we are forced to work
with truncated model wave functions $\Psi_0$.
If the matrix element of
a nuclear operator $\cO$ is calculated
using model wave functions,
then
$\mbox{$<\!\Psi^f|\cO|\Psi^i\!>$} \neq
<\!\Psi_0^f|\cO|\Psi_0^i\!>$.
This deviation represents the core-polarization effect.
The core polarization effects
need to be carefully sorted out
before one can identify
the exchange currents effects.
Despite this non-trivial aspect,
there is growing evidence
that supports the \CPT \,derivation of exchange currents.
The best example is
the nuclear axial-charge operator $A_0$.
Warburton \etal's
systematic analyses\cite{war91,wt94}
of the first-forbidden $\beta$ transitions
indicate that the ratio of
the exchange-current contribution
to the 1-body contribution is
$\delta_{\rm{mec}}\equiv
\langle A^0({\rm mec})\rangle/
\langle\!A^0(1\hyphen{\rm body})\rangle
\!=\!0.6 \sim 0.8$.
(The semi-empirical method used
in these analyses largely eliminates ambiguities
due to the core-polariation effects.)
The leading-order \CPT \,term,
\ie the soft-pion exchange
term,$\!$\cite{kdr78,rho91}
can explain the bulk of $\delta_{\rm{mec}}$,
and the next-order
\CPT\,term\cite{pmr93,ptk94}
gives an additional $\sim$10\% enhancement,
bringing the theoretical value close to
the empirical value.

It is informative to compare
the above results with those obtained
in the conventional
meson-exchange approach.$\!$\cite{krt92,tow92}
Using the ``hard-pion formalism"
in conjunction with the lagrangian
that engenders the phenemenological
$N$-$N$ interactions,
Towner\cite{tow92} finds
that the pion-exchange contribution
is reduced significantly
by the phenomenological form factors,
but the reduction is largely compensated
by heavy-meson pair graphs.
The net result is:
$\delta_{\rm{mec}}^{\rm Towner}
\sim\delta_{\rm{mec}}^{\rm CPhT}$.
In fact, the former is slightly larger,
but this small difference is qualitatively
understood as follows.
The largest heavy-meson pair contributions
come from $\sigma$ and $\omega$ mesons,
and the $\sigma$-meson contribution
can be effectively rewritten
as the $1$-body term with the nucleon mass
replaced by an effective mass.$\!$\cite{dt87}
Thus the phenomenological $\sigma$-meson
plays a role similar to the BR scaling.$\!$\cite{br91}
Meanwhile, in \CPT,
the BR scaling is attributable
to multi-fermion terms
which have higher chiral orders than
those appearing in the next-to-leading-order
calculation of Park \etal$\!$\cite{pmr93,ptk94}
Then, we should qualitatively expect
$\delta_{\rm{mec}}$
obtained by Park \etal \,
to be somewhat smaller than
$\delta_{\rm{mec}}^{\rm Towner}$.
The above example demonstrates
the usefulness of CPT\,
in organizing complicated exchange-current
contributions in a systematic manner.

For the two-nucleon systems
we can obtain exact solutions
for $H_N\Psi= E\Psi$,
avoiding thereby the core-polarization problem.
The A=2 systems therefore
provide a clean case for checking the validity of
the standard calculational framework
based on the nucleonic Schr\"{o}dinger equation
supplemented with the exchange currents.
A beautiful test is found
in radiative capture of a thermal neutron
by a proton:
$n+p \ra d+\gamma$.
The observed capture rate for this process is
$\sigma_{\rm exp}=334.2 \pm 0.5$mb,
which is $\sim$10\% larger than the IA prediction
$\sigma_{IA}=302.5\pm4.0$mb.
According to Riska and Brown,$\!$\cite{rb72}
the one-pion exchange current derived
from the low-energy theorem
can account for $\sim$70\% of the
missing capture rate.
Recognizing that this contribution represents
the leading order term in \CPT,
it is of great interest to examine
what the next-order term will do.
Park \etal's recent calculation\cite{pmr95}
that includes
the next-to-leading order terms gives
$\sigma=334\pm2$mb,
in perfect agreement with experiment.
(Another impressive success of
the exchange current calculations
based on the low-energy theorem
is known for the
$e+d\ra e+p+n$ reaction,
see \eg \,Ref.\cite{fm89}.)

Our last topic is neutrino reactions on
the deuteron.
The recent developments
in the solar neutrino problem
have further enhanced the importance of
the MSW effect
as a possible mechanism to explain
the observed energy dependence
of the solar neutrino deficit.$\!$\cite{sch95}
The SNO heavy-water
\v{C}erenkov counter\cite{ardetal87}
can provide crucial information on this issue
because of its capability
to register the charged-
and neutral-current reactions
simultaneously but separately.
The SNO is also expected to be highly useful
for studying supernova neutrinos.
The neutrino-deuteron reactions
relevant to the SNO are:
$\nu + d \ra \nu' + n + p$,
$\bar{\nu} + d \ra \bar{\nu} + n + p$,
$\nu_e + d \ra e^- + p + p$ and
$\bar \nu_e + d \ra e^+ + n + n$.
Obviously, one needs reliable estimates
of the cross sections
for these reactions
to extract useful astrophysical information
from SNO data.
The above discussion indicates that
one can have enough confidence
in the calculational framework
that uses the nucleonic Schr\"{o}dinger equation
with realistic $N$-$N$ interactions
supplemented with the exchange currents.
Although one may eventually be able to
obtain all the ingredients from \CPT,
it is reasonable to use phenomenological input.
There have been several calculations
of this type\cite{tkk90,kk92},
and the best available estimates (in our opinion)
have been given by Kohyama \etal$\!$\cite{kk92}

\end{document}